# Visualizing Orbital Content of Electronic Bands in Anisotropic 2D Semiconducting ReSe$_2$

Byoung Ki Choi,[1] Søren Ulstrup,[2,3] Surani M. Gunasekera,[4] Jiho Kim,[5] Soo Yeon Lim,[6] Luca Moreschini,[3] Ji Seop Oh,[3,7,8] Seung-Hyun Chun,[9] Chris Jozwiak,[3] Aaron Bostwick,[3] Eli Rotenberg,[3] Hyeonsik Cheong,[6] In-Whan Lyo,[5] Marcin Mucha-Kruczynski,[4] and Young Jun Chang,[1,*]

[1]*Department of Physics, University of Seoul, Seoul 02504, Republic of Korea*
[2]*Department of Physics and Astronomy, Aarhus University, Denmark, 8000 Aarhus C, Denmark*
[3]*Advanced Light Source (ALS), E. O. Lawrence Berkeley National Laboratory, Berkeley, California 94720, USA*
[4]*Centre for Nanoscience and Nanotechnology and Department of Physics, University of Bath, Bath BA2 7AY, United Kingdom*
[5]*Department of Physics, Yonsei University, Seoul, 03722, Republic of Korea*
[6]*Department of Physics, Sogang University, Seoul, 04107, Republic of Korea*
[7]*Center for Correlated Electron Systems, Institute for Basic Science (IBS), Seoul 08826, Republic of Korea*
[8]*Department of Physics and Astronomy, Seoul National University, Seoul 08826, Republic of Korea*
[9]*Department of Physics, Sejong University, Seoul 05006, Republic of Korea*

[*]e-mail: yjchang@uos.ac.kr

**ABSTRACT**

Many properties of layered materials change as they are thinned from their bulk forms down to single layers, with examples including indirect-to-direct band gap transition in 2H semiconducting transition metal dichalcogenides as well as thickness-dependent changes in the valence band structure in post-transition metal monochalcogenides and black phosphorus. Here, we use angle-resolved photoemission spectroscopy to study the electronic band structure of monolayer $ReSe_2$, a semiconductor with a distorted 1T structure and in-plane anisotropy. By changing the polarization of incoming photons, we demonstrate that for $ReSe_2$, in contrast to the 2H materials, the out-of-plane transition metal $d_z^2$ and chalcogen $p_z$ orbitals do not contribute significantly to the top of the valence band which explains the reported weak changes in the electronic structure of this compound as a function of layer number. We estimate a band gap of 1.7 eV in pristine $ReSe_2$ using scanning tunneling spectroscopy and explore the implications on the gap following surface-doping with potassium. A lower bound of 1.4 eV is estimated for the gap in the fully doped case, suggesting that doping-dependent many-body effects significantly affect the electronic properties of $ReSe_2$. Our results, supported by density functional theory calculations, provide insight into the mechanisms behind polarization-dependent optical properties of rhenium dichalcogenides and highlight their place amongst two-dimensional crystals.

Among the plethora of two-dimensional (2D) materials entering the spotlight following the discovery of graphene,[1] transition metal dichalcogenides (TMDs) attract huge interest as potential building blocks for innovative electronic and optoelectronic applications.[2–4] TMDs, with their diverse and intriguing properties, such as semiconductivity, superconductivity, charge-density wave order, ferroelectricity, and ferromagnetism,[2,4–10] open distinct device platforms as components in stacks of 2D crystals. As bulk materials are thinned down to monolayer (ML) forms to be used in such stacks, some of their properties might change due to quantum confinement effects. For example, semiconducting TMDs, such as $MX_2$ (M=Mo, W and X=S, Se), exhibit an indirect electronic band gap in the bulk but a direct one in the ML limit, making the latter promising candidate materials for optoelectronics. This indirect-to-direct transition as a function of decreasing layer number has been confirmed with various experimental methods, such as photoluminescence,[4] ellipsometry,[11] and angle-resolved photoelectron spectroscopy (ARPES).[12] The strong layer-dependent evolution of the electronic structure indicates a significant interlayer coupling between individual layers.

At the same time, rhenium compounds ($ReS_2$ and $ReSe_2$) are expected to undergo very small variations in the electronic band structure as a function of thickness due to weaker interlayer coupling than in other semiconducting TMDs.[13–15] As opposed to the more common 1T structure (Fig. 1a), $ReSe_2$ is characterized by a 1T′ structure with triclinic symmetry (space group: P1),[16] induced by a Jahn-Teller-like structural distortion and seen as a zig-zag Re chain structure (red lines) in Fig. 1b. Such strong distortion induces anisotropy in the layer plane, resulting in, amongst others, strong linear dichroism.[15,17] It also leads to larger interlayer spacing and introduces a large band gap into the electronic dispersion which would otherwise feature a half-filled metallic ground state.[18] In contrast to semiconducting TMDs with 2H structure, the electronic band structure of $ReSe_2$ and its orbital composition have not been well understood especially in the ML limit. ReSe$_2$ shows nearly layer-independent photoluminescence signal, suggesting that a direct band gap is preserved for all thicknesses.[19] Standard density functional theory (DFT) calculations predict an anisotropic valence band maximum (VBM) with heavy effective mass located slightly away from the Γ point. When considering electron correlation effects, the VBM is rather uniform around the Γ point, supporting the presence of a direct band gap in the ML.[20] In case of bulk ReSe$_2$, previous electronic structure studies found anisotropic 3D bulk bands with ~0.1 eV modulation along the k$_z$ direction driven by the interlayer interactions.[21] In the similar compound ReS$_2$, the 3D bulk band structure with in-plane anisotropy shifts the VBM away from the Γ point along k$_z$, while ML and bilayer band structures indicate a direct band gap character with an increased effective hole mass in the ML.[22,23] However, a direct experimental study of the electronic band structure of ML ReSe$_2$, especially its orbital characteristics, from which one can understand the highly anisotropic carrier transport, optical, and vibrational properties, as well as interlayer interactions across the van der Waals spacing, is lacking.[13,24–26]

In this report, we study the electronic band structure of ML ReSe$_2$ which we grew epitaxially on a bilayer graphene (BLG)/SiC (0001) substrate by using molecular beam epitaxy (MBE). From ARPES measurements, we obtained the ML ReSe$_2$ valence band structure and used polarization-resolved data as well as DFT calculations to elucidate the orbital contributions of Re to the valence band. We show that the out-of-plane transition metal (Re) $d_{z^2}$ and chalcogen (Se) $p_z$ orbitals do not significantly contribute to states forming the VBM but rather to those in subbands located at higher binding energies ($E_{bin}$) in the range of -2 ~ -2.5 eV. This is in contrast not only to 2H TMDs but also many other 2D materials, including post-transition metal monochalcogenides and black phosphorus, in which parts of the valence band are formed by states extending in the out-of-plane direction and are therefore strongly affected by interlayer coupling.[27–33] As a result, the shape of the valence band in these materials depends sensitively on layer number, leading for example to indirect-to-direct band gap transitions.[34] Our results provide a direct explanation of the relatively weak band structure changes in ReSe$_2$ as a function of film thickness and emphasize the distinctness of rhenium compounds amongst not only the TMDs but also two-dimensional crystals in general. We also estimate the size of the band gap in our ReSe$_2$ samples in both pristine and surface-doped cases, obtaining a gap of 1.7 eV in the pristine sample, while a lower bound of 1.4 eV is found for fully doped ReSe$_2$. The implications of these observations on many-body effects are discussed.

**RESULTS AND DISCUSSION**

Figure 1b schematically shows the ML ReSe$_2$ on a BLG substrate. The crystal structure consists of a distorted hexagonal plane of Re atoms sandwiched between two corrugated Se layers, forming a 1T′ structure with triclinic symmetry. The strong lattice distortion expands the unit cell (black dashed parallelograms) twice along both crystallographic directions in the plane, leading to the formation of zigzag chains of Re atoms (red lines), clearly distinct from the undistorted 1T structure in Fig. 1a. The direction along the Re chains is roughly aligned with the zigzag direction of the underlying BLG on 6H-SiC (0001), while the minimum lattice mismatch is estimated to be 1.6% for the match between 3 ReSe$_2$ unit cells and 8 graphene unit cells. Figure 1c shows the reflection high-energy electron diffraction (RHEED) pattern of the substrate. The BLG serves as an ideal substrate for fabricating van der Waals heterostructures and for studying their electronic structures due to the chemical inertness.[5,12,35] The RHEED images in Fig. 1d show clear streaky lines indicating a well-ordered epitaxial growth of ML ReSe$_2$ on top of the graphene lattice. Compared to the other ML 2H[12,36–38] or 1T phases[39] showing (1 × 1) RHEED patterns, additional peaks (green arrows) indicate an increase of unit cell size due to strong lattice distortion of the 1T′ phase, similar to the other 1T′ TMDs.[40] However, while in other TMDs the unit cell is often doubled due to a change of periodicity along only one of the in-plane directions, in Re dichalcogenides the unit cell area is approximately quadrupled as both in-plane primitive lattice vectors increase. Referring to the in-plane lattice constant of BLG (a = 2.46 Å), the in-plane lattice constants of ReSe$_2$ are calculated to be ~6.65 ± 0.1 Å, similar to those of the bulk (6.72 Å ($a_1$) and 6.61 Å ($a_2$)).[16]

In Fig. 1e, the Raman spectrum of our film shows sharp peaks corresponding to the reported peak positions (red circles) in ReSe$_2$ flakes,[26] indicating the highly crystalline state of the film, comparable to the flakes. ReSe$_2$ has a large number of Raman peaks, because the reduced crystal symmetry in the 1T′ distorted structure splits the Raman peaks which are degenerate in the high-crystal symmetry 1T structures. Figure 1f shows a large-scale STM image with ReSe$_2$ film coverage of 0.6 ML on BLG (the corresponding RHEED pattern is shown in the inset of Fig. 1h). In Fig. 1g, the height of ReSe$_2$ islands (~6.65 Å) is consistent with the c-axis lattice constant of the bulk (6.72 Å). In the zoomed-in STM image (Fig. 1h), the ML ReSe$_2$ shows well-ordered crystalline phases along with noticeable stripe patterns, which indicate the anisotropic nature of the 1T′ phase. Therefore, the ML ReSe$_2$ films are epitaxially grown on BLG with well-ordered crystalline and topographic quality. We also note that the randomly distributed lumps are probably selenium oxide or related amorphous components that originated during the selenium decapping procedure after the *ex-situ* sample transfer.

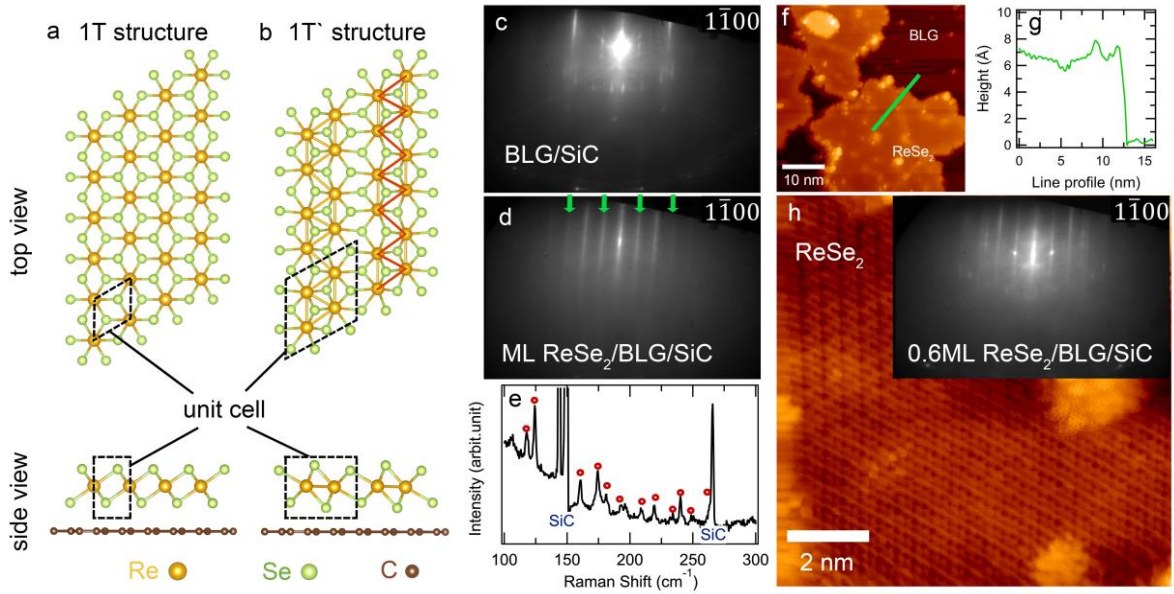

Figure 1. Morphology of ML ReSe$_2$ on BLG. (a,b) Top- and side-view schematics of ML ReSe$_2$ with 1T and 1T′ structures with graphene substrate added in the side view. The yellow, green, and brown balls represent the Re, Se, and C atoms, respectively and the red zigzag line shows the rhenium chains. (c,d) RHEED images of BLG and ML ReSe$_2$, reflecting epitaxial growth of ReSe$_2$ on BLG. The distortion in the 1T′ ReSe$_2$ induces additional peaks (green arrows). (e) Raman spectrum of ML ReSe$_2$ on BLG with sharp film peaks marked with red circles and two SiC substrate peaks. (f) Topographic STM image ($V_{Sample}$ = −2.0 V, $I_T$ = 50 pA) of 0.6 ML ReSe$_2$ grown on BLG. (g) Line profile along the green line in f. (h) Magnified STM image (−2.0 V, 50 pA) in the same region shows stripe patterns due to the anisotropic Re chain structure (Inset: RHEED image of corresponding 0.6 ML ReSe$_2$ sample).

Figure 2 presents ARPES constant-energy (CE) maps of 1 ML ReSe$_2$ grown on BLG at various $E_{bin}$ obtained with p- and s-polarized light. Signal from the π bands of BLG dominates near the graphene's Brillouin zone (BZ) corner (marked as K$_G$) for both polarizations. Band structure of the 1 ML ReSe$_2$ becomes evident only after saturating the graphene band. Because of the doubling of the in-plane lattice constants, the area of the BZ of ReSe$_2$ is reduced to a quarter (black hexagons) as compared to the 1T phase (green dashed hexagon), which also explains the additional (2 × 2) RHEED patterns. The two-dimensionality of our film is evidenced by the photon-energy-independent band structure (see Fig. S1), to be contrasted with the dependence on photon energy observed across the 3D BZ of bulk ReSe$_2$.[21,41] In addition, on a hexagonal graphene substrate, the ReSe$_2$ islands can nucleate to form three energetically equivalent domains, each with the anisotropic crystal axis rotated by approximately 120° with respect to each other, similar to ReS$_2$.[42] Since the ReSe$_2$ island size is about a few tens of nanometers in size (much smaller than the photon beam size of 20~50 micrometer), as shown in Fig. 1f, the ARPES intensity in the maps represents an average signal from all three anisotropic domains. This is why, although the BZ for a single crystal ML ReSe$_2$ should be a slightly distorted regular hexagon, here we can work with an effective perfectly hexagonal BZ. Because of that, we use the labels K and M to denote any of the corners and centers of a side of a hexagonal BZ, respectively, rather than distinguish between different directions as would be necessary for a single domain. Finally, this explains why the CE maps reveal nearly isotropic electronic states around principal points (K and M), as shown in Fig. 2b,e (see Fig. S2), as opposed to the anisotropic valence bands of the bulk single crystals and flakes of ReSe$_2$ and ReS$_2$.[21,23,43] We note that similar multi-domain effects have been observed in ARPES maps of ML 1T′ WTe$_2$ on BLG.[40]

The CE map at the Fermi level ($E_F$) shows only background intensity at the Γ point and signatures of graphene bands around all six K$_G$ (Fig. 2a,d), indicating that the $E_F$ is located inside the band gap of ML ReSe$_2$. For $E_{bin}$ of -1.1 eV, the electronic states emerge forming a nearly isotropic valence band at every Γ point for both p- and s-polarizations (Fig. 2b,e). From the intensity distribution in the CE map, we find that the VBM is centered at the Γ point. Looking at energies below $E_{bin}$ = -1.1 eV, we can see enlarged valence band contours centered at the Γ points as shown for $E_{bin}$ of -1.3 eV in Fig. 2c,f.

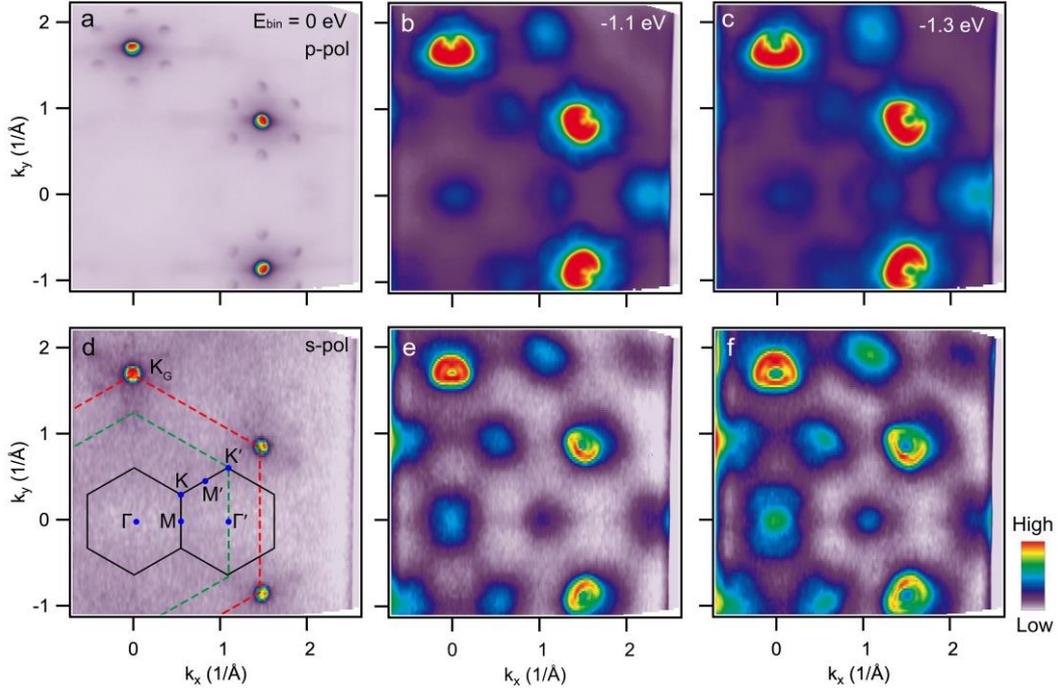

Figure 2. Constant energy ARPES maps of 1T′ ML ReSe$_2$. (a-f) CE maps of ML ReSe$_2$ with three selected $E_{bin}$ (= 0, -1.1, -1.3 eV) with p-(a-c) and s-polarized (d-f) light ($h\nu$ = 96 eV, T = 90 K). The solid lines indicate the first and second Brillouin zones (BZs) of ML ReSe$_2$ with 1T′ structure. The green and red dashed lines correspond to the BZs of the undistorted 1T structure and BLG, respectively.

In Fig. 3a-d, the valence band dispersions along both Γ - M and Γ – K directions also show symmetric features centered at the Γ point, for both p- and s-polarization cases. In Fig. 3e, however, our DFT calculation (technical details of our computational method can be found in the density functional theory calculations part of the Methods section) of the ML ReSe$_2$ band structure shows an anisotropic energy surface within the BZ (black hexagon), with a small dip at the Γ point. When we take into account the multi-domain effect by overlapping band contours rotated by 120° (red, yellow, and blue surfaces), as shown in Fig. 3f for the highest valence band, the calculated band structure becomes essentially 3-fold symmetric. The presence of a small dip at the Γ point is not fully in agreement with the centered isotropic feature in our CE maps, possibly due to insufficient treatment of the electron-electron interactions.[20]

The appearance of both the ML ReSe$_2$ VBM and BLG neutrality point in our ARPES data allows us to analyze the mutual alignment of the band structures of the two materials, shown as a schematic diagram in Fig. 3g. The neutrality point of BLG is positioned at the energy $E_N$ = -0.33 eV as seen in Fig. 3b,d, which is consistent with the range of -0.24 ~ -0.32 eV for a BLG on *n*-doped SiC.[44,45] Using the results of work function measurements for BLG on SiC,[44] we estimate the BLG neutrality point to be 4.54 eV below the vacuum energy ($E_{vac}$). From our data fitted with gaussian functions (see Fig. S3 in the Supporting Information), the VBM of ReSe$_2$ is situated -1.1 ± 0.05 eV below $E_F$ and 0.77 eV below the BLG neutrality point, see Fig. 3b and d. This places the ReSe$_2$ VBM 5.31 eV below $E_{vac}$. At the same time, our DFT calculation predicts the VBM position in pristine ReSe$_2$ as 5.24 eV below $E_{vac}$ (compared with an experimental value of 5.6 eV for bulk ReSe$_2$),[46] suggesting a minor Schottky barrier of ≲0.1 eV, similar to that deduced for MoS$_2$ on graphene.[47] Such a small barrier value implies a weak charge interaction between the 1 ML ReSe$_2$ and BLG. In addition, we performed temperature dependent ARPES measurements between 80 K and 270 K, as shown in Fig. S4. During the temperature cycle, we observe a ~50 meV energy shift, which likely results from the complex temperature dependent carrier densities in both ReSe$_2$ and the SiC substrate.[48] This small shift implies that the temperature-induced change of carrier density does not significantly affect the electronic structure of ML ReSe$_2$ on BLG. We also note that all the ARPES measurements were performed at 80 K, so that the temperature effect is minimal for understanding the ARPES results.

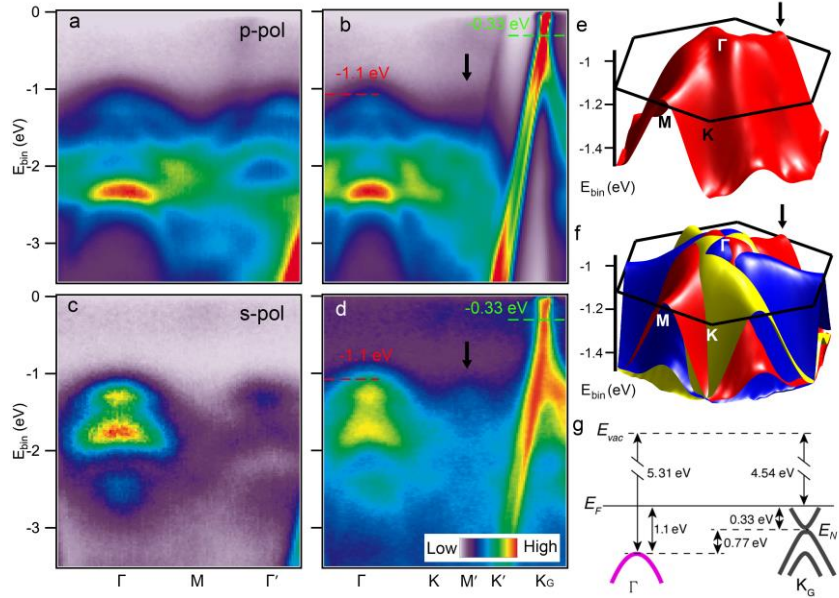

Figure 3. Valence band structure of ML ReSe$_2$. (a-d) ARPES spectra along Γ-M-Γ′ and Γ-K-M′ for ML ReSe$_2$ with the p-(a,b) and s-polarization (c,d). (e,f) Anisotropic energy surface within the first BZ (black hexagons) for a single domain (red, e) and for three domains related effectively by 120° rotations (red, yellow, blue, f). The black arrows in (b) and (d) indicate the intensity close to the M′ point from electronic states of one of the domains as marked with the same arrows in e and f for the red surface. (g) Band alignment of pristine ML ReSe$_2$ on BLG.

The orbital characteristics of the band structure is of great importance for understanding interlayer coupling and optical absorption processes.[49,50] To identify the orbital contributions to the ReSe$_2$ valence band, we compared ARPES spectra measured with p- and s-polarized light. The changes in the photocurrent intensity for the two polarizations are mostly due to the matrix element effects during the photoemission process, which encode the orbital character of the bands.[40,51,52] Within the dipole approximation, the electronic state in the crystal needs to be odd (even) with respect to reflection in the scattering plane in order to lead to a nonzero photocurrent for s-(p-) polarization.[52,53] In Fig. 4a,b, the valence band spectra indeed show contrasting intensity distributions depending on light polarization. The distribution with p-(s-) polarized light is intense at -2 ~ -2.5 eV (-1 ~ -2 eV), respectively. While the presence of multiple domains makes it difficult to relate the orbitals to the scattering plane, two of them, the rhenium $d_z^2$ and selenium $p_z$ orbitals, have circular symmetry in the layer plane and hence couple to p-polarized photons. This suggests that the $d_z^2$ and $p_z$ orbital states are concentrated deeper in the valence band rather than close to the VBM. We corroborate this with our calculations of orbital projections of the valence band states, Fig. 4c,d. These show that the $d_z^2$ and $p_z$ orbital states are concentrated near the lower part of the valence bands, while the rest of the d-orbitals ($d_{x^2-y^2}$ and $t_{2g}$) form states in the upper part of the valence band, consistent with the previously suggested electronic structure for bulk ReSe$_2$.[18,46] We present orbital projections along the whole momentum range shown in Fig. 3a-d, which show qualitative agreement with the ARPES intensity, in the Supporting Information, Fig. S5. Our observation is also in agreement with the suggestion that the orbital makeup of valence band states in bulk ReSe$_2$ is responsible for the observed negative pressure coefficients of its excitonic transitions.[54] In particular, the energy of the transition assigned to the Z point of the bulk BZ, which is projected on the Γ point in the two-dimensional BZ of ML, undergoes the weakest changes with increasing pressure because the corresponding VBM states are formed mostly by orbitals less extended out of the plane than $d_z^2$ which are less affected by the applied pressure.

It is worth noting that in many layered materials including semiconducting 2H TMDs, post-transition metal monochalcogenides, and black phosphorus, the top valence band states are made of orbitals extending significantly out of the plane and responsible for interlayer interactions (mainly $d_z^2$ and $p_z$). The lack of nearest neighbor layers in the case of ML crystals lowers the energy of such states while parts of the electronic dispersion made of in-plane orbitals remain largely unaffected. In the 2H semiconducting dichalcogenides, the transition metal $d_{z^2}$ and chalcogen $p_z$ orbitals form the top-most valence-band states around Γ while the in-plane orbitals contribute to the states around K. With increasing number of layers, interlayer coupling splits the energies of the states at Γ (which would be degenerate in the case of a stack of non-interacting layers), so that some of them increase in energy and overcome the local VBM at K. This is accompanied by similar changes in the bottom conduction band where electronic states at K, forming the conduction band minimum in the ML, are overtaken in energy by states closer to Γ, resulting in a transition from a direct band gap in the ML to an indirect one in the bulk.[12,34,55] Similar shift of the top valence band states dominated by contributions of $p_z$ orbitals leads to a significant decrease of electronic band gaps in the bulk as compared to the ML in monochalcogenides, such as, InSe, GaSe and GaS.[27–33] However, in this group, an upwards shift of the states at Γ in thicker crystals leads to an indirect-to-direct band gap from ML to bulk (with conduction band minimum always located at Γ irrespectively of layer number). Black phosphorus, instead, remains a direct gap semiconductor for all thicknesses, with in-plane anisotropy leading to interesting photo-optoelectronic properties.[56,57] As a result, the orbital composition of the valence band states, visualized in Fig. 4, makes rhenium dichalcogenides distinguished amongst the layered materials. This orbital composition is a consequence of the symmetry of the (undistorted) 1T structure and, more importantly, an additional valence electron of Re as compared to Mo or W forming 2H semiconducting TMDs. As a result of the latter, undistorted 1T ReSe$_2$ would be a metal with several $d_{x^2-y^2}$- and $d_{xy}$-character bands close to the $E_F$.[18] These $d$ orbitals lie in the layer plane and thus are strongly affected by any in-plane distortions which modify Re-Re couplings. The large distortion driving the shift from the 1T to 1T′ structure lowers the energy of the mentioned bands across the whole new, smaller BZ. The affected bands move down in energy but remain above other occupied bands and so form the top valence band of the now semiconducting ReSe$_2$. Ultimately, this explains the weak crystal-thickness-dependence of the electronic band gap in ReSe$_2$ and possibly also in ReS$_2$.[13]

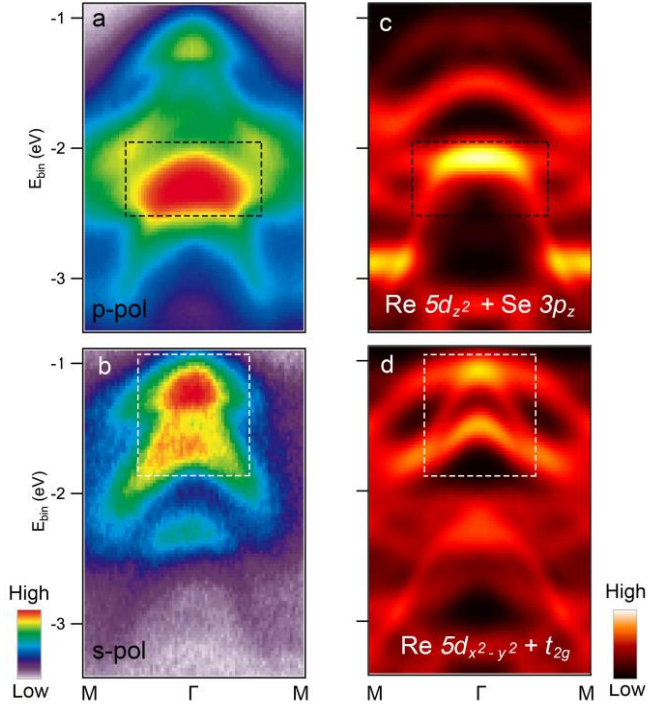

Figure 4. Orbital-selective valence band structure of ML ReSe$_2$. (a,b) Valence band dispersion along M-Γ-M for *p*- (a) and *s*-polarized light (b). (c,d) Calculated density of states projections on the Re 5$d_z^2$ and Se 3$p_z$ (c) or the remaining Re *d* orbitals (d). The black and white dashed boxes represent the intense distributions for the ARPES maps with each polarization, which are well accounted for by the calculated intensity profile for $d_z^2 + p_z$ or the sum of all other *d* orbitals, respectively.

A detailed comparison of the ARPES data and our band structure calculation is presented in Fig. 5a-d, where the theoretical dispersion is plotted using red, yellow, and blue lines corresponding to the three different domains with colors to be compared to the surfaces in Fig. 3f. Overall, the calculated bands overlap well with the ARPES features. The relative rotations of the three domains lead to three distinctive energy dispersions. However, notice that for each of the surfaces in Fig. 3e,f, the dispersion along one of the M-Γ-M directions is relatively flat while it crosses a steep hill in the perpendicular K-Γ-K direction. The former corresponds to the direction perpendicular and the latter parallel to the Re chains.[21] Because of the relative rotations, each of the domains contributes a dispersive feature along one of the Γ-K directions along its K-M′-K′ segment and we have marked one of these with black arrows in the ARPES maps in Fig. 3b,d and 5b,d, as well as in Fig. 3e,f (backfolded into the first BZ). Together, the flat and dispersive behavior of the electronic bands implies that charges are localized (delocalized) along the direction perpendicular (parallel) direction to the Re chains, as reflected by nearly flat (dispersive) regions of the calculated top valence band drawn in yellow (red) in Fig. 5.

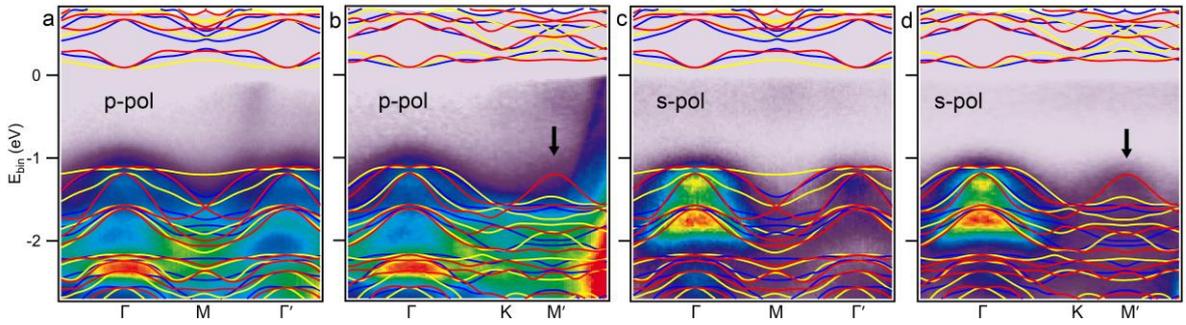

Figure 5. Comparison of DFT and ARPES bands of ML ReSe$_2$. (a-d) Comparison of the ARPES data with p-/s-polarizations and our band structure calculations for the three rotated domains (red, yellow, and blue). The black arrows indicate the distinct dispersive states contributed by one of the domains (red).

To support our conclusion of only weak changes in the valence band structure of ReSe$_2$ as a function of thickness, we further compare its valence band structure to that of a bilayer (2 ML). In Fig. 6a,b, we compare (2 × 2) RHEED patterns for 1 and 2 ML, where the main change in the 2 ML case is an increased broadening of the streaks. As presented in Fig. 6c,d, the valence band structure of 2 ML does not exhibit an abrupt change of its features compared to the 1 ML, but displays a relatively small energy shift of ~0.2 eV for both the VBM and the intense subbands at higher $E_{bin}$ around Γ, as indicated with dashed lines. Our experimental data is consistent with our DFT calculations which indicates little change in the valence band between 1 ML and 2 ML ReSe$_2$. This energy shift is consistent with the small change (0.22 eV) of the optical band gap between 1 ML and 2 ML ReSe$_2$ flakes.[19] Both the absence of an abrupt band structure change and the small band shift are in line with the weak thickness-dependence of the shape of the valence band in ReSe$_2$. We note that the increased broadening of the ARPES intensity in the 2 ML data is probably due to structural disorder in the second layer, similar to the broadened RHEED pattern.

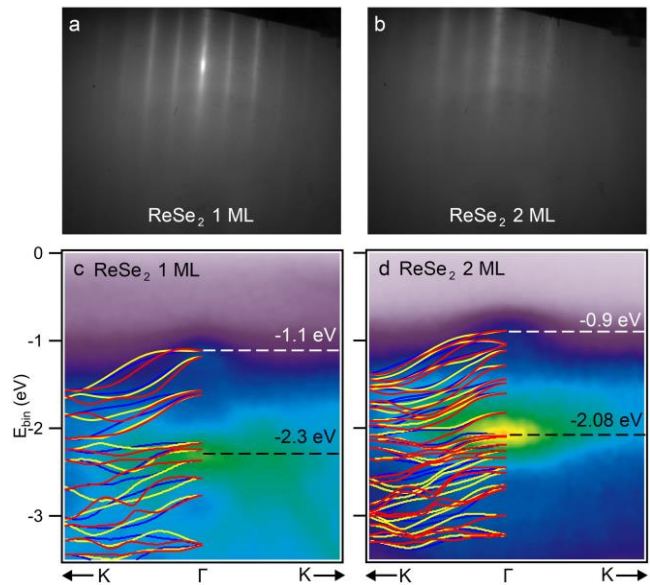

Figure 6. Comparison of 1 ML and 2 ML ReSe$_2$. (a,b) RHEED images of 1 ML and 2 ML ReSe$_2$. The enhanced disorder and multi-domain effect may increase broadening in the 2 ML sample. (c,d) ARPES bands with DFT calculations along K-Γ-K for 1 ML (c) and 2 ML ReSe$_2$(d). White and black dashed lines indicate the energy positions of the VBM and an intense band towards higher $E_{bin}$ at the Γ point.

As a non-trivial semiconducting TMD, it is essential to estimate the band gap of ML ReSe$_2$. Since ARPES only measures the occupied electronic states, we deposited potassium onto the sample surface during measurements in order to electron dope the structure and thereby attempt to shift the conduction bands below the $E_F$. The doping effect saturates at large potassium concentrations at which potassium simply accumulates on the surface without further electron donation (see Fig. S6).[12,58,59] Figure 7a-c show valence band cuts and a maximum energy shift of 0.31 eV (from -1.1 eV to -1.41 eV) during the potassium-dosage until a coverage of 1.2 ML was reached (we define 1 ML coverage by tracking the appearance of an interface peak of potassium 3$p$ using X-ray photoemission spectroscopy, see Fig. S6). The evolution of energy dispersion curves at the Γ point reveals a gradual energy shift of the valence bands, as shown in Fig. 7a-d. However, the surface-doping *via* potassium not only induces an energy shift of the ReSe$_2$ bands (0.31 eV), but also shifts the underlying BLG bands down by 0.2 eV (see Fig. S7). Following the model of BLG from [ref 60], we calculated the 2D carrier density change in the BLG from the pristine state (1.6 × 10$^{13}$ cm$^{-2}$, $E_N$= -0.33 eV) to the case of a 1.2 ML potassium coverage (4.1 × 10$^{13}$ cm$^{-2}$, $E_N$= -0.53 eV).[60] Assuming comparable carrier doping for both the ReSe$_2$ and BLG, we can then estimate the actual doping density of 2~3 × 10$^{13}$ cm$^{-2}$ for 1 ML potassium coverage, equivalent to 0.04~0.06 electrons per potassium atom.

Ultimately, we do not observe intensity from the ReSe$_2$ conduction band states near the $E_F$ during potassium doping, as doping beyond 1.2 ML does not shift the bands of either ReSe$_2$ or BLG any further, but enhances a broad background signal, probably due to the potassium adatoms accumulating on the surface. Our DFT calculation shows quite flat conduction bands, as seen in Fig. 5a-d, and hence a considerable electronic density of states which requires a large number of electrons to populate – one of the possible reasons why we do not fill the conduction band states. For this reason, we can use the VBM position of -1.41 ± 0.05 eV as a lower bound on the magnitude of the electronic band gap, by assuming that the $E_F$ is close to the minimum energy of such flat conduction bands. Our lower bound estimate is similar to the optical band gap in ML ReSe$_2$ (1.32 ~ 1.47 eV),[14,19] but somewhat smaller than the gap of ~1.7 eV, which we obtained from scanning tunneling spectroscopy (STS) measurements on our ReSe$_2$/BLG heterostructure, as shown in Fig. 7e. The big difference between the optical and STS gap is explained by the large exciton binding energy (*i.e.* 460 – 680 meV) resulting from strong many-body interactions.[61] The uncertainty of the STS gap may also arise from the tip-induced band bending owing to poor screening of electric fields at a semiconductor surface.[62] On the other hand, while both ARPES and STS measure the quasi-particle band gap, the difference between the ARPES lower bound and the STS value likely occurs due to the fact that the increased carrier density significantly alters the dielectric screening and thus the Coulomb interaction in the system, which renormalizes the band gap.[63–65] This renormalization of the gap can mainly be described in terms of a rigid band shift. In ML WSe$_2$, the observed decrease of the gap with respect to the value for the undoped structure was of the order of 0.5 eV for doping of ~$10^{13}$ cm$^{-2}$,[66,67] which is similar to the estimated additional density provided by potassium in our case, supporting the notion that at 1.2 ML potassium coverage we might be close to occupying the conduction band states. Note that in the ML limit, the reduction of interlayer screening increases the exciton binding energy while the dielectric screening from the underlying substrates becomes increasingly important.[62] In addition, the strong interfacial electronic interaction also induces in-gap states and effectively reduces the STS gap in the case of semiconducting TMD-metal interfaces, as demonstrated for example for Au substrates.[68–71] Finally, further investigation of hetero-interfaces between ReSe$_2$ and other materials are of great interest due to possible electronic states related to moiré potentials and interface states.[61,68]

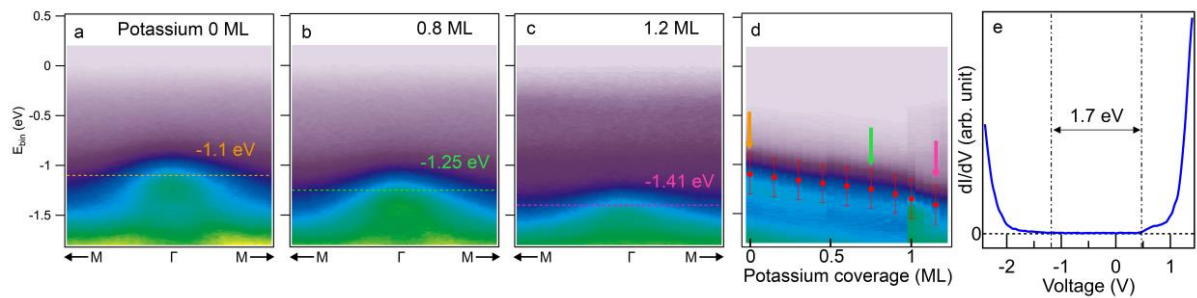

Figure 7. Band gap estimation of ML ReSe$_2$. (a-c) Evolution of valence bands around the Γ point with three potassium coverages of 0 (a), 0.8 (b) and 1.2 ML (c). (d) Shift of valence band maximum with potassium-coverage at Γ. Red circles indicate fitted band positions. The three arrows indicate the potassium coverage of a-c, respectively. (e) STS spectrum ($V_{Sample}$ = 1.5 V,

$I_T$ = 100 pA) of pristine ML ReSe$_2$ with estimated band gap of 1.7 eV.

**CONCLUSIONS**

In summary, we have successfully grown ML ReSe$_2$ epitaxially on BLG. We confirmed the single-crystalline properties of the ML-thick nanoscale ReSe$_2$ islands by STM and Raman spectroscopy and resolved their valence band structure using ARPES. We also confirmed that, in contrast to many other two-dimensional crystals including the semiconducting 2H phases, *i.e.* MoX$_2$ and WX$_2$ (X=S, Se), monochalcogenides InSe, GaSe and GaS, and black phosphorus, in ReSe$_2$ the $d_z^2$ transition metal orbital contributes only weakly to the electronic states at the top of the valence band at Γ. Such orbital character as well as the lattice distortion significantly suppress interlayer interaction in this material. Our direct visualization of the orbital content of the valence band of ReSe$_2$ with ARPES highlights the intriguing properties of this anisotropic 2D layered semiconductor with potential for future applications in electronic and optoelectronic devices. Finally, we provide a lower bound of 1.41 eV on the electronic band gap in doped ML ReSe$_2$ following potassium deposition and also an estimate of 1.7 eV from STS in the pristine state. The different gap sizes highlight the significance of doping induced many-body effects in ReSe$_2$.

**METHODS**

**Molecular beam epitaxy.** ReSe$_2$ films were grown by using a home-built MBE chamber with a base pressure of $1 \times 10^{-10}$ Torr. We used 6H-SiC (001) single crystal substrates, supplied by the Crystal Bank at Pusan National University. We annealed 6H-SiC substrates at 1300°C in 2 min for 3 times in ultra-high vacuum (UHV) condition followed by the formation of BLG layer on the Si-terminated surface of SiC. We co-evaporated Re (99.97%) and Se (99.999%) by using an e-beam evaporator and an effusion cell, respectively. During film growth, we maintained a substrate temperature of 250°C for 10 min for 1 ML thickness and then annealed the sample at 420°C for 30 min. In case of 2 ML sample, we extend the deposition duration accordingly. We carried out *in situ* RHEED measurements with a high voltage of 18 kV. For both STM and ARPES measurements, we covered the samples with an amorphous selenium layer at room temperature to protect the pristine surface from air exposure after the film growth. The samples were then annealed at 480 K in UHV to remove the selenium capping layer.

**Raman spectroscopy.** A 532 nm (2.33 eV) DPSS laser was used as an excitation source for Raman measurements. The laser beam was focused on the sample with a ~ 1 µm diameter by a 50 × objective lens (N.A.=0.8) which also collected the scattered light from the sample. The scattered light was dispersed with a Jobin-Yvon Horiba iHR550 spectrometer (2400 grooves/mm) and was detected with a charge-coupled-device (CCD) using liquid nitrogen for cooling. We kept the laser power below 0.1 mW to avoid damages from local heating. The spectral resolution is ~ 1 cm$^{-1}$.

**Scanning tunneling microscopy and spectroscopy.** The experiments were performed in an UHV chamber ($3 \times 10^{-11}$ mbar) by using a variable-temperature STM (OMICRON GmbH) equipped with home-built software and electronics. Electrochemically etched and Ar-sputtered tungsten tips were used. All STM and STS measurements were conducted at 79 K. The bias voltages ($V_{sample}$) stated in the topographic images were applied to the sample. For the STS spectrum, each $I(V)$ spectrum consists of 1024 data points and was Gaussian smoothed maintaining thermally limited energy resolution of $\Delta E = 28$ meV. $dI/dV$ spectrum was acquired by numerical differentiation of the $I$-$V$ sweep.

**Angle resolved photoemission spectroscopy.** ARPES measurements were performed in the micro-ARPES end-station (base pressure of ~$3 \times 10^{-11}$ Torr) at the MAESTRO facility at beamline 7.0.2 at the Advanced Light Source, Lawrence Berkeley National Laboratory, using p- or s-polarized photon. Depending on the polarization, the photoemission matrix element causes the intensity of the measured spectral features to vary strongly.[72,73] The ARPES system was equipped with a Scienta R4000 electron analyzer. The lateral size of the synchrotron beam was estimated to be between 30 and 50 μm. The sample temperature was kept at ~80 K during all the ARPES measurement. The total energy resolution was 20 meV at $h\nu = 96$ eV and calibrated with a polycrystalline gold film. A series of measurements was made with various photon energies in the range of 60 – 150 eV. Potassium-deposition was carried out by evaporation of potassium on the sample surfaces using a commercial SAES getter source mounted in the analysis chamber such that doping is performed in an optimum sample measurement position without moving the sample. Measurements and doping experiments were carried out at 90 K. The amount of potassium deposited can be estimated from the potassium 3*p* core level spectra (Supplementary Fig. S6a). The potassium 3*p* peak initially appears at $E_{bin}$ = ~18 eV and grows in intensity as the initial potassium ML is completed. When the second potassium layer starts to grow, a chemically shifted potassium 3*p* peak emerges at $E_{bin}$ = ~19 eV.[58,59] Re 4*f* core-level spectra shifts accordingly with the energy shift of 0.317 eV upon reaching a potassium coverage of 1.2 ML (Supplementary Fig. S6).

**Density functional theory calculations.** DFT calculations used the Quantum ESPRESSO[74,75] package and pseudopotentials from the PSlibrary[76]. For the band structure surfaces and cuts in Fig. 3 and Fig. 5, we used fully relativistic local density approximation (LDA) pseudopotentials, with the Perdew-Zunger[77] parametrization of the exchange correlation energy. For the orbital projections along M-Γ-M in Fig. 4, we used scalar-relativistic generalized gradient approximation (GGA) pseudopotentials (so that wave function projections onto atomic states classified only by orbital angular momentum could be conveniently obtained), with the Perdew-Burke-Ernzerhof[78] parametrization of the exchange correlation energy. In all cases, the valence of Re was taken as 15, and no van der Waals corrections were included. Results obtained using LDA/GGA scalar- and fully-relativistic pseudopotentials are compared in Ref. 26 and 41.[26,41] Atomic coordinates were taken from the Chemical Database Service[79] and relaxed to obtain forces less than 0.006 eV/Å. Kinetic energy cutoffs were typically, 60 Ry (816 eV) and Monkhorst-Pack[80] k-point meshes of at least $8 \times 8 \times 1$ were used. To include the effect of different domains for Fig. 3f, we produced cuts along the three non-equivalent K-Γ-K (M-Γ-M) directions and plotted them together in such a way that the position of the left-most Γ point corresponds to the Γ point at (0,0) for all the cuts.

**ASSOCIATED CONTENT**

Supporting Information

The Supporting Information is available free of charge on the ACS Publications website at DOI: Further characterization on the photon energy-dependent ARPES data, symmetric band structure along the three M-Γ-M or K-Γ-K directions, the valence band maximum and core level spectra during surface electron-doping, the temperature dependence, the orbital character of valence bands of ML ReSe$_2$.

The authors declare no competing financial interests.

**AUTHOR INFORMATION**

**Corresponding Authors**

*E-mail: yjchang@uos.ac.kr.

**Author Contributions**

B.K.C., and S.U. contributed equally. Y.J.C. conceived the experiments. B.K.C., S.-H.C., and Y.J.C. prepared thin films. J.K. and I.-W.L. carried out STM experiments. S.Y.L. and H.C. performed Raman measurements. B.K.C., S.U., L.M., J.O., C.J., A.B., E.R., and Y.J.C. performed ARPES measurements. S.M.G and M.M.-K. carried out theoretical calculations. B.K.C., S.U., M.M-K., and Y.J.C. prepared the manuscript. All authors discussed the results and commented on the manuscript.

**ACKNOWLEDGMENTS**

This work was supported by the National Research Foundation (NRF) grants funded by the Korean government (No. NRF-2020R1A2C200373211, 2019K1A3A7A09033389, and 2019R1A2C3006189) and made use of the Balena High Performance Computing (HPC) Service at the University of Bath. S.U. acknowledges financial support from the VILLUM FONDEN (Grant No. 15375). The Advanced Light Source is supported by the Director, Office of Science, Office of Basic Energy Sciences, of the U.S. Department of Energy under Contract No. DE-AC02-05CH11231. S.M.G. acknowledges support from the University of Bath and U.K. Engineering and Physical Sciences Research Council (EPSRC) through the Centre for Doctoral Training in Condensed Matter Physics, Grant EP/L015544/1EPSRC (UK). M.M.-K. was supported through the International Funding Scheme of the University of Bath. This work was supported by IBS-R009-D1. Authors thank B.L. Chittari and J. Kim for fruitful discussion.

**Supporting Information**

# Visualizing Orbital Content of Electronic Bands in Anisotropic 2D Semiconducting ReSe$_2$

Byoung Ki Choi,[1] Søren Ulstrup,[2,3] Surani M. Gunasekera,[4] Jiho Kim,[5] Soo Yeon Lim,[6] Luca Moreschini,[3] Ji Seop Oh,[3,7,8] Seung-Hyun Chun,[9] Chris Jozwiak,[3] Aaron Bostwick,[3] Eli Rotenberg,[3] Hyeonsik Cheong,[6] In-Whan Lyo,[5] Marcin Mucha-Kruczynski,[4] and Young Jun Chang,[1,*]

[1]*Department of Physics, University of Seoul, Seoul 02504, Republic of Korea*
[2]*Department of Physics and Astronomy, Aarhus University, Denmark, 8000 Aarhus C, Denmark*
[3]*Advanced Light Source (ALS), E. O. Lawrence Berkeley National Laboratory, Berkeley, California 94720, USA*
[4]*Centre for Nanoscience and Nanotechnology and Department of Physics, University of Bath, Bath BA2 7AY, United Kingdom*
[5]*Department of Physics, Yonsei University, Seoul, 03722, Republic of Korea*
[6]*Department of Physics, Sogang University, Seoul, 04107, Republic of Korea*
[7]*Center for Correlated Electron Systems, Institute for Basic Science (IBS), Seoul 08826, Republic of Korea*
[8]*Department of Physics and Astronomy, Seoul National University, Seoul 08826, Republic of Korea*
[9]*Department of Physics, Sejong University, Seoul 05006, Republic of Korea*

[*]e-mail: yjchang@uos.ac.kr

# 1. 2D character of valence band structure in ML ReSe$_2$

Figure S1 shows photon energy ($h\nu$)-dependence of the valence band structure in ML ReSe$_2$ for both p- and s-polarized light configurations. Energy and momentum positions of the bands do not change for different $h\nu$ values, although their relative intensities vary due to different cross-sections or geometric effects. The lack of dispersion with $h\nu$ is consistent with the 2D nature of the electronic structure in the single-layer limit.

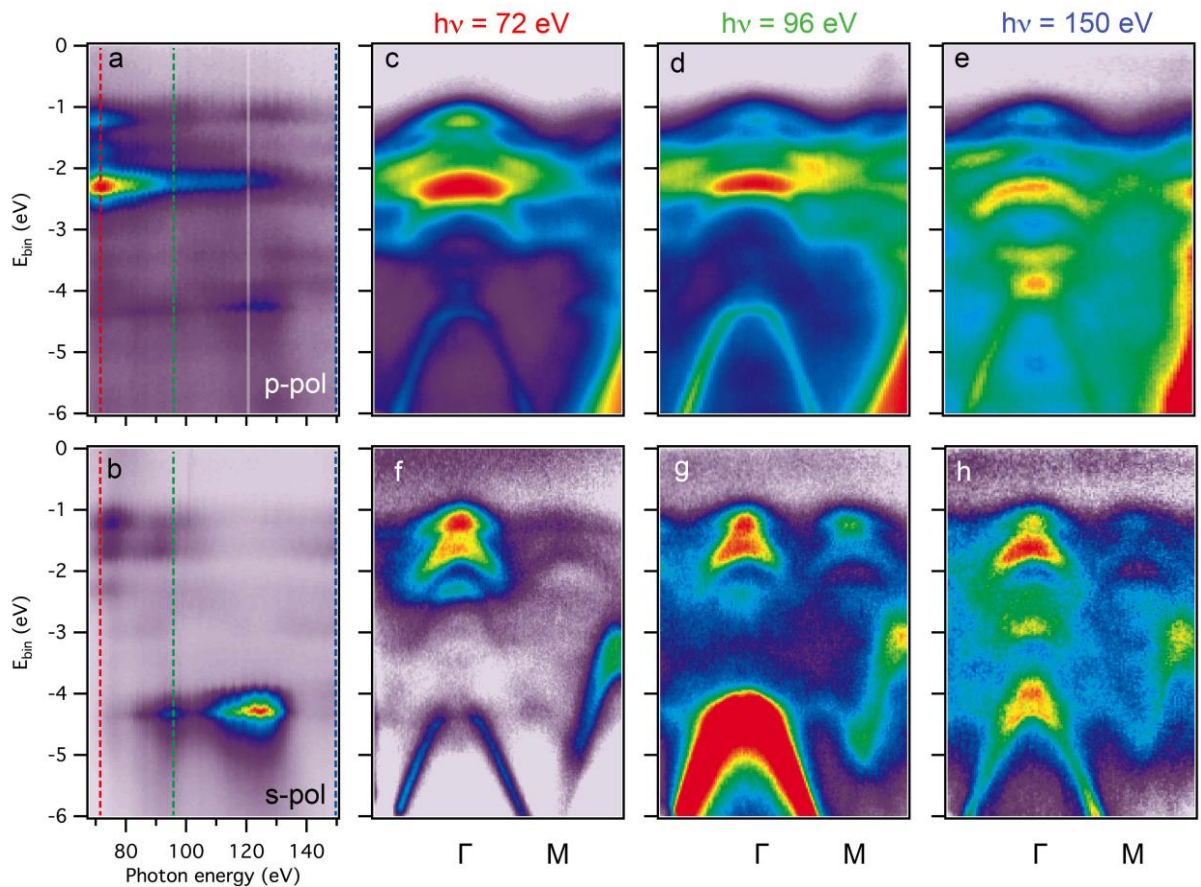

Figure S1. 2D character of valence band structure in ML ReSe$_2$. (a,b) Photon energy($h\nu$)-dependence of valence band cuts at Γ with p- (a) and s-polarized light (b). (c-e) Valence bands along Γ-M for $h\nu$ = 72 (c), 96 (d), 150 eV (e) using p-polarization. (f-h) Valence bands along Γ-M for $h\nu$ = 72 (f), 96 (g), 150 eV (h) using s-polarization. The 2D electronic structure of the ML ReSe$_2$ is confirmed by the lack of dispersion with $h\nu$.

## 2. Symmetric band structure of ML ReSe$_2$ along three equivalent axes

Figure S2 shows symmetric valence band structure of ML ReSe$_2$ along the principal axes ($\Gamma$-M$_i$ and $\Gamma$-K$_i$ for i=1,2,3). The band cuts display similar features along each of the $\Gamma$-M$_i$ or $\Gamma$-K$_i$ directions, showing their effective equivalence.

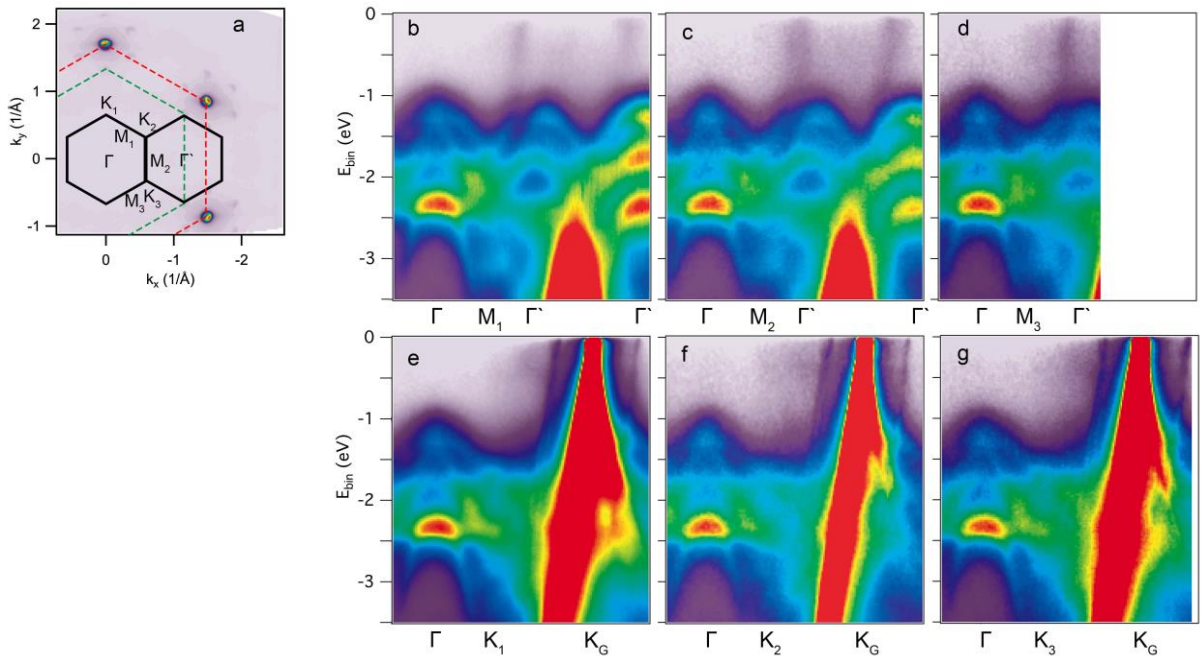

Figure S2. Symmetric valence band structure of ML ReSe$_2$ along the principal axes. (a) Fermi surface map of ML ReSe$_2$. (b-g) Band cuts along $\Gamma$-M$_1$ (b), $\Gamma$-M$_2$ (c), $\Gamma$-M$_3$ (d), $\Gamma$-K$_1$ (e), $\Gamma$-K$_2$ (f), and $\Gamma$-K$_3$ (g) directions. (All spectra her were obtained with the p-polarization and $h\nu = 96$ eV.)

## 3. Valence band maximum analysis during potassium deposition

Figure S3 presents an energy distribution curve (EDC) analysis of a polycrystalline Au spectrum and the ReSe$_2$ valence band maximum (VBM) during potassium deposition. The EDC extracted from the Au spectrum is fitted by a Fermi-Dirac distribution. VBM fits of ReSe$_2$ consist of two components (VB1 and VB2) that are described by gaussian functions.

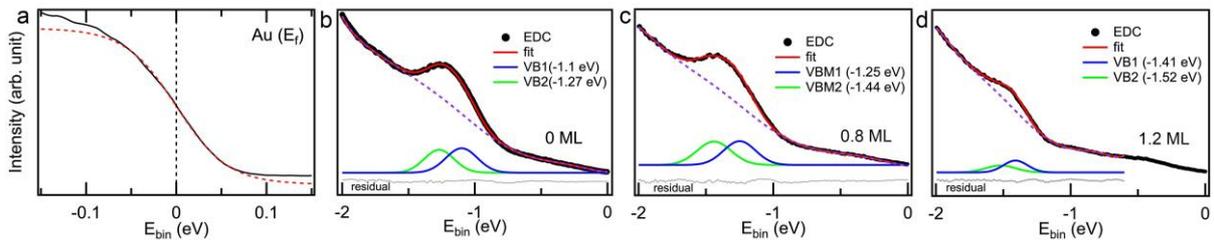

Figure S3. VBM shift during potassium deposition. (a) EDC around the Fermi edge taken on polycrystalline Au in order to calibrate the energy scale. (b-d) VBM fits of ML ReSe$_2$ spectra for a potassium coverage of 0 ML (b), 0.8 ML (c), and 1.2 ML (d).

## 4. Temperature dependence of valence band structure in ML ReSe$_2$ and BLG

Figure S4 presents a comparison between the ARPES spectra taken at sample temperatures of 80 K and 270 K. During the temperature cycle, we observe ~50 meV energy shift of the ReSe$_2$ VBM, which is a relatively small change. This small shift implies that a temperature-induced change of carrier density does not significantly alter the electronic structure of ML ReSe$_2$ on BLG. We also note that all the measurements were performed at ~90 K, so that the temperature effect is minimal for understanding the rest of ARPES results.

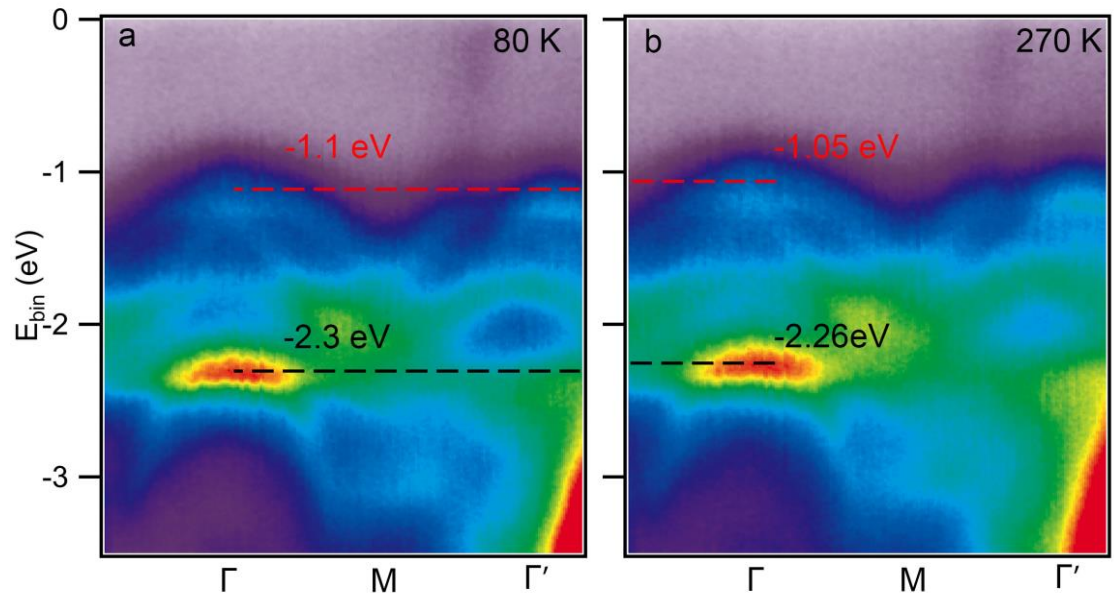

Figure S4. Temperature dependence of valence band structure in ML ReSe$_2$. (a,b) ARPES spectra for ML ReSe$_2$ along Γ-M-Γ′ at 80 K (a) and 270 K (b).

## 5. Orbital character of valence bands

Figure S5 presents the orbital-projected valence band calculation for a wide momentum range. These orbital projected bands show qualitative agreement with the ARPES spectra in Fig. 3a-d.

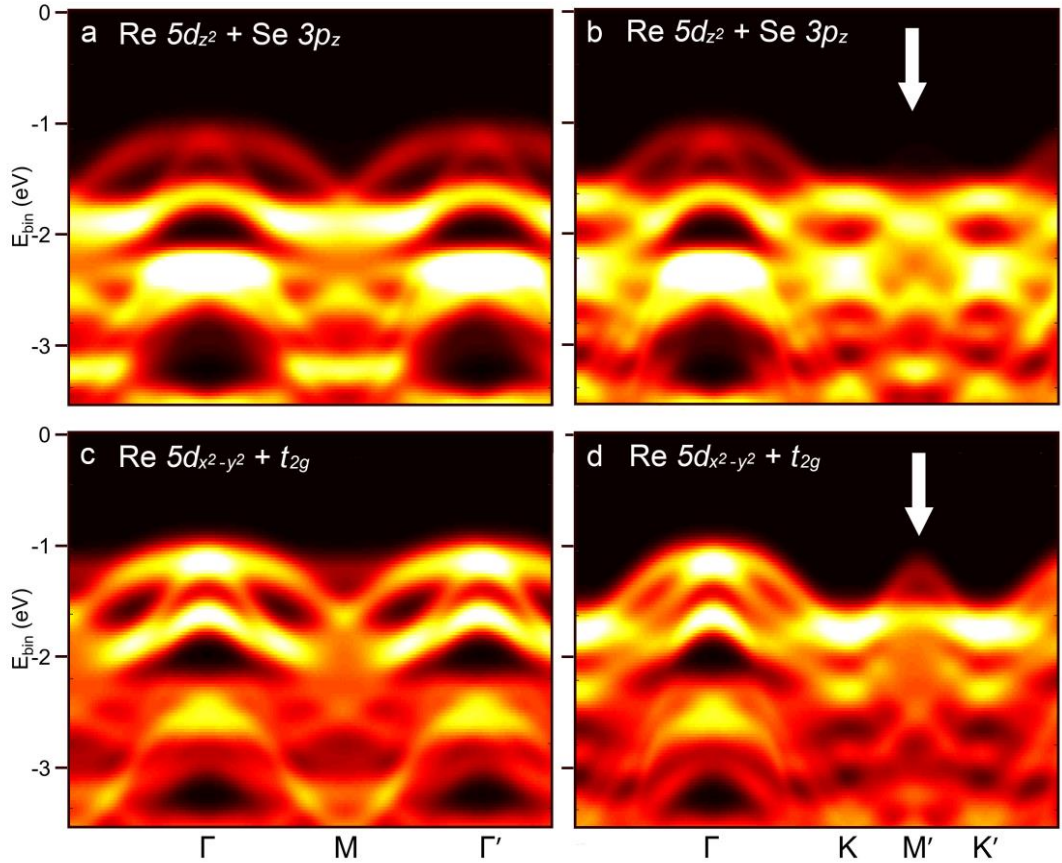

Figure S5. Orbital-projected valence band density of states in ML ReSe$_2$. (a,b) Contribution of Re $5d_z^2$ and Se $3p_z$ orbitals along Γ-M-Γ′ (a) and Γ-K-M′ (b). (c,d) Contribution of Re $d_{x2-y2}$ and $t_{2g}$ orbitals along Γ-M-Γ′ (c) and Γ-K-M′ (d). The white arrows in (b) and (d) indicate the intensity close to the M′ point from electronic states of one of the domains also marked in Fig. 3.

## 6. Surface electron doping *via* potassium deposition

Figure S6 shows X-ray photoemission spectroscopy (XPS) data of potassium (3*p*) and rhenium (4*f*) collected during surface electron doping *via* potassium deposition. The potassium 3*p* spectra exhibit a systematic intensity increase with potassium coverage. When the second potassium layer forms, an interface peak emerges on the left of the main surface peak. We define potassium density at which the interface peak emerges as one monolayer (1 ML) coverage. During the potassium deposition, the rhenium 4*f* spectra show an energy shift of -0.32 eV as seen in Fig. S6d, which is consistent with the VBM shift (-0.3 eV) discussed in the main manuscript. It is noted that the small shoulders (Re-Se) on the right side of main peaks are probably caused by the defect phase formed during Se-decapping. These shoulders do not shift during the potassium deposition.

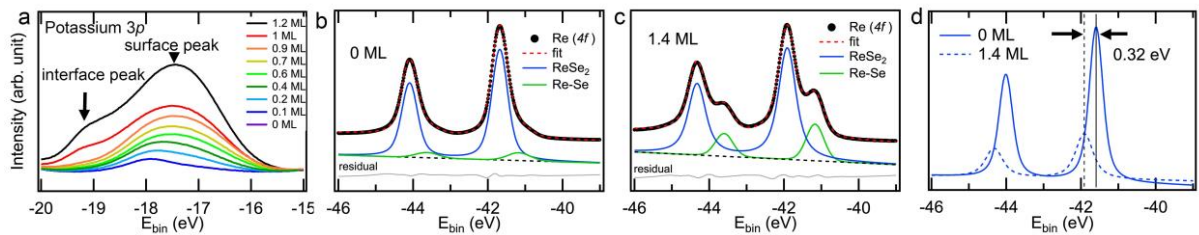

Figure S6. Surface electron doping *via* potassium deposition. (a) XPS data of potassium 3*p*. (b-d) XPS data of rhenium 4*f* at a coverage of 0 ML (b) and 1.4 ML (c). (d) Comparison of the main peaks (ReSe$_2$) for the two coverages.

# 7. Doping dependence of valence band structure in ML ReSe$_2$ and BLG

Figure S7 presents the valence bands of ReSe$_2$ and bilayer graphene for different surface potassium doping levels in order to compare the respective band alignments. In the pristine sample, the charge neutrality point of the BLG is given by $E_N = -0.33$ eV below the Fermi level. In comparison, VBM of ReSe$_2$ is located at $E_{VBM} = -1.1$ eV below the Fermi level. At 1.2 ML potassium coverage, the $E_N$ shifts to -0.53 eV below the Fermi level while the VBM moves to $E_{VBM} = -1.41$ eV. During the doping, the 2D carrier density of BLG changes from $1.6 \times 10^{13}$ cm$^{-2}$ to $4.1 \times 10^{13}$ cm$^{-2}$ (we estimate these carrier densities by using BLG density of states as provided by a standard tight-binding model), which corresponds to $2.1 \times 10^{13}$ cm$^{-2}$ per ML potassium coverage. Since this carrier density only accounts for the doping of the BLG, the total contributing carrier density from the potassium should also include the doping of ML ReSe$_2$.

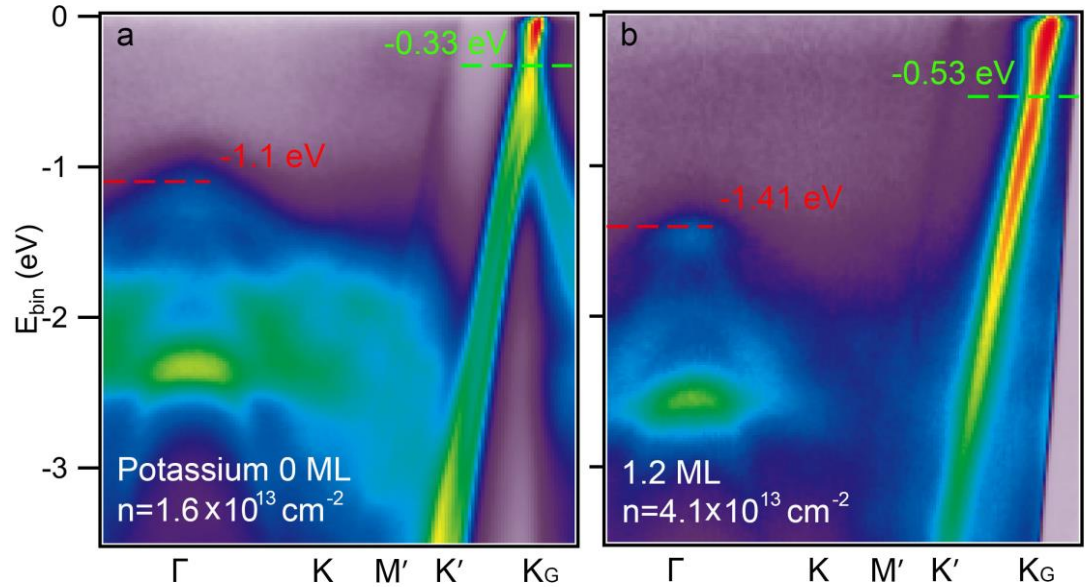

Figure S7. Doping dependence of valence band structure in ML ReSe$_2$ and BLG. (a,b) ARPES spectra for ML ReSe$_2$ along Γ-K-M′ for potassium coverage of 0 ML (a) and 1.2 ML (b). Red and green dashed lines indicate the VBM of ML ReSe$_2$ and the neutrality point ($E_N$) of BLG, respectively.